\documentclass{article}
\input{amssym}
\begin{document}
\title{The Generalized Classical Time-Space}
\date{}
\author{Mehdi Nadjafikhah\thanks{Department of Mathematics, Iran University of Science and
Technology, Narmak-16, Tehran, Iran. e-mail:
m\_nadjafikhah@iust.ac.ir.}\and Seyed-Mehdi Mousavi\thanks{e-mail:
mahdi\_mousavi@mathdep.iust.ac.ir }}
\maketitle
\begin{abstract}
The newest model for space-time is based on sub-Riemannian
geometry. In this paper, we use a combination of Lorentzian and
sub-Riemannian geometry, the suggest a new model which likes to
its ancestors, but with the most efficient in application. In
continuation, we try to show a new connection which calls
generalized connection, and prove some its properties.
\end{abstract}

\section{Introduction}
In physics, Newtonian and Einstein theories are used as
non-quantomic space-time theories. In Einstein's general
relativity, a Lorentzian manifold is also considered as a
space-time coordinates. This manifold contains some infirmations
of gravity and electromagnetism. In fact, a space-time is a time
process while some forces are acting. As usual, in physics,
relativity theory is considered as a principle axiom. Newtonian
theory is noticed as a limit case of relativity theory and it is
used for justification of relativity. Of course, in spite of
relativity, there are another theories in this field, just as Elis
restricted covariance, Hartel-Howking classic limit, Logonov
relative theory, and EPS axioms offered by Ehler, Pirani, and
Shield. Through these theories, EPS axioms are most popular among
physicists.

%%%%%%%%%%%%%%%%%%%%%%%%%%%%%%%%%%%%%
\section{The generalized classical time-space}
\paragraph{Definition 1.} An ordered quadric set
$( M , \Omega , \big< . , . \big> , \nabla )$ is called a {\it
generalized space-time} when satisfies in following conditions
\begin{itemize}
\item[(1)] $M$ is a smooth and connected manifold of dimension
$m=n+1\geq2$,
\item[(2)] $\Omega$ is a non-zero differential 1-form: for every
$p\in M$, then $\Omega_p\neq0$,
\item[(3)] if we define
\begin{eqnarray*}
{\rm Ann}\,\Omega&:=& \{v\in TM : \Omega(v)=0\} \\
\Gamma({\rm Ann}\,\Omega) &:=& \{V\in\Gamma(TM) :
V_p\in\mathop{\rm Ann}\,\Omega, \forall p\in M\}
\end{eqnarray*}
then
\begin{eqnarray*} \big<.,.\big> :
\Gamma(\mathop{\rm Ann}\,\Omega)\times\Gamma(\mathop{\rm
Ann}\,\Omega)\longrightarrow
C^\infty(M)\,:\hspace{0.5cm}(V,W)\mapsto \big<V,W\big>
\end{eqnarray*}
defines an inner product on $\Gamma(\mathop{Ann}\,\Omega)$ that is
symmetric but is not positive-definite necessarily.
\item[(4)] $\nabla$ is a linear connection, named as {\it generalized
connection}, so that it satisfies in the following two conditions
\begin{itemize}
\item[(i)] $\nabla\Omega=0$ that is, for each $X\in\Gamma(TM)$,
then $\nabla_X\Omega=0$,
\item[(ii)] $\nabla\big<.,.\big>$=0, that is for every
$X\in\Gamma(TM)$ and $V,W\in\Gamma({\rm Ann}\,\Omega)$ then
\begin{eqnarray*}
X\big<V,W\big>=\big<\nabla_XV,W\big>+\big<V,\nabla_XW\big>.
\end{eqnarray*}
\end{itemize}
\end{itemize}
\paragraph{Definition 2.} From {\it field of observers} - in
abbreviation FO - we mean a vector field $z\in\Gamma(TM)$ such
that
$\Omega(z)=1$.\\

Let ${\cal Z}(M)$ be the set of all FOs. We define a field of
automorphisms in the following form
\begin{equation}\label{1}
P_z := v-\Omega(v)z\ \ \ \ \forall v \in\ \  TM.
\end{equation}
\paragraph{Theorem 1.} In every generalized space-time
$(M,\Omega,\big<.,.\big>,\nabla)$, we have
\begin{equation}\label{2}
\Omega\circ{\rm Tor}=d\Omega
\end{equation}
where, ${\rm Tor}$ is torsion of the generalized connection.
\paragraph{Corollary 1.} If the generalized connection be
symmetric, then we have $d\Omega=0$.
\paragraph{Definition 3.} Let $z\in{\cal Z}(M)$ be an FO.
Induced {\it gravity field} of $\nabla$ in $z$ is following vector
field
\begin{equation}\label{3}
{\cal{G}}=\nabla_zz,
\end{equation}
also, induced {\it Coriolis field} of $\nabla$ in $z$ is a field
of non-symmetric covariant 2-form $\omega=\frac{1}{2}{\rm
rot}\,z$, so that for each $V,W \in \Gamma(\mathop{an}\,\Omega)$
we have
\begin{eqnarray}\label{4}
\omega(V,W)=\frac{1}{2}\Big(\big<\nabla_Vz,W\big>-\big<V,\nabla_Wz\big>\Big).
\end{eqnarray}
%%%%%%%%%%%%%%%%%%%%%%%%%%%%%%%%%%%%%
\section{Existence of the generalized connection}
\paragraph{Theorem 2.} Let $(M,\Omega,\big<.,.\big>,\nabla)$ be a
generalized space-time and $z\in{\cal Z}(M)$ be an FO with gravity
field ${\cal G}$ and Coriolis field $\omega$. Then for every
$X,Y\in\Gamma(TM)$ and every $V\in\Gamma(\mathop{Ann}\,\Omega)$,
$\nabla$ satisfies in the following relation
\begin{eqnarray}
2\big<P_z(\nabla_XY),V\big> &=& X\big<P_z(Y),V\big>+Y\big<P_z(X),V\big> \nonumber \\
&&-V\big<P_z(X),P_z(Y)\big>+2(\Omega(X)\Omega(Y)\big<{\cal G},V\big>\nonumber \\
&&+\Omega(X)\omega(P_z(Y),V)+\Omega(Y)\omega(P_z(X),V)) \\
&&+\Omega(X)(\big<A(z,P_z(X)),V\big>-\big<A(z,V),P_z(Y)\big>)\nonumber \\
&&-\Omega(Y)(\big<A(z,P_z(X)),V\big>-\big<A(z,V),P_z(X)\big>)\nonumber \\
&&+\big<A(P_z(X),P_z(Y)),V\big>-\big<A(P_z(Y),V),P_z(X)\big>\nonumber \\
&&-\big<A(P_z(X),V),P_z(Y)\big>\nonumber,\label{5}
\end{eqnarray}
in which, $A(X,Y)=\nabla_XY-\nabla_YX $.\\

\noindent{\it Proof:} Since $\nabla\Omega=0 $ and $\nabla g=0$ we
use of the following recursive relations
\begin{eqnarray}
V\big<P_z(X),P_z(Y)\big> &=& \big<\nabla_VP_z(X),P_z(Y)\big>+\big<P_z(X),\nabla_VP_z(Y)\big>,\label{6}\\
P_z(X)\big<P_z(Y),V\big> &=& \big<\nabla_{P_z(X)}P_z(Y),V\big>+\big<P_z(Y),\nabla_{P_z(X)}V\big>,\label{7}\\
P_z(Y)\big<V,P_z(X)\big> &=&
\big<\nabla_{P_z(Y)}V,P_z(x)\big>+\big<V,\nabla_{P_z(Y)}P_z(X)\big>.\label{8}
\end{eqnarray}
By computing (\ref{7})+(\ref{8})-(\ref{6}) we have
\begin{eqnarray}
&&\hspace{-2cm} \big<\nabla_{P_z(X)}P_z(Y)+\nabla_{P_z(Y)}P_z(X),V\big> = \nonumber\\
&=& P_z(X)\big<P_z(Y),V\big>+P_z(Y)\big<V,P_z(X)\big>\nonumber \\
&=& \big<\nabla_{P_z(X)}P_z(Y),V\big>+\big<\nabla_{P_z(Y)}P_z(X),V\big>\nonumber \\
&&
+\big<A(P_z(x),P_z(Y)),V\big>+2\Omega(X)\big<\nabla_zP_z(Y),V\big>.\label{9}
\end{eqnarray}
On the other hand, by the definitions of $P_z$ and $A$ we have
\begin{eqnarray}
2\big<\nabla_XP_z(Y),V\big>\!\!\!&=&\!\!\!\! 2\big<\nabla_{P_z(X)}P_z(Y),V\big>\!\!
+\!\!2\Omega(X)\big<\nabla_zP_z(Y),V\big>\nonumber \\
&&-V\big<P_z(X),P_z(Y)\big>-\big<A(P_z(Y),V),P_z(X)\big> \nonumber\\
&&-\big<A(P_z(X),V),P_z(Y)\big>.\label{10}
\end{eqnarray}
By setting (\ref{9}) in (\ref{10}) we conclude that
\begin{eqnarray}
2\big<\nabla_XP_z(Y),V\big> &=& P_z(X)\big<P_z(Y),V\big>+P_z(Y)\big<V,P_z(X)\big> \nonumber \\
& & -V\big(P_z(X),P_z(Y)) + \big<A(P_z(Y),V),P_z(X)\big> \label{11}\\
& & -\big<A(P_z(X),V),P_z(Y)\big>+\big<A(P_z(X),P_z(Y)),V\big> \nonumber\\
& & + 2 \Omega(X)\big<\nabla_zP_z(Y),V\big>. \nonumber
\end{eqnarray}
Also if we set the values $P_z(X)$ and $P_z(Y)$ in the right hand
of relation (\ref{11}), we have
\begin{eqnarray}
2\big<\nabla_XP_z(Y),V\big> &=& \Omega(X)\big<\nabla_zP_z(Y),V\big>-\Omega(X)\big<P_z(Y),\nabla_zV\big> \nonumber \\
&&-\Omega(Y)\big<\nabla_zV,P_z(X)\big>-\Omega(Y)\big<V,\nabla_{P_z(X)}z\big> \label{12}\\
&& -\Omega(Y)\big<V,A(z,P_z(X))\big>\nonumber +\{{\rm Koszul}\},
\nonumber
\end{eqnarray}
where
\begin{eqnarray}
\{{\rm Koszul}\} &=& X\big<P_z(Y),V\big>+Y\big<V,P_z(X)\big> \nonumber \\
& & -V\big<P_z(X),P_z(Y)\big> + \big<A(P_z(X),P_z(Y)),V\big>\label{13} \\
& & -\big<A(P_z(Y),V)\big>-\big<A(P_z(X),V),P_z(Y)\big>.\nonumber
\end{eqnarray}
Now by considering the following relation
$$\nabla_X(\Omega(Y)z)=\Omega(\nabla_XY)z+\Omega(Y)(\Omega(X)\nabla_zz+\nabla_{P_z(X)}z)$$
we have
\begin{eqnarray}
P_z(\nabla_XY) &=& \nabla_XY-\Omega(\nabla_XY) \nonumber \\
&=& \nabla_X(\Omega(Y)z)+\nabla_XP_z(Y)-\Omega(\nabla_XY)z \label{14}\\
&=&\Omega(X)\Omega(Y){\cal{G}}+\Omega(Y)\nabla_{P_z(X)}z+\nabla_XP_z(Y).
\nonumber
\end{eqnarray}
Finally, by setting (\ref{13}) in (\ref{12}) proof will be
complete. \hfill $\diamond$
\paragraph{Theorem 3.} Suppose that the space-time
$(M,\Omega,\big<.,.\big>)$ has first three conditions of
generalized space-times and ${\cal{D}}(\Omega,\big<.,.\big>)$ be
the set of all generalized connections. Then for a fixed FO, $z$,
the map
$$D^z:{\cal{D}}(\Omega,\big<.,.\big>)\longrightarrow
\Gamma(\mathop{\rm Ann}\,\Omega)\times\wedge^2(\mathop{\rm
Ann}\,\Omega) \times\wedge^2(TM,\mathop{\rm Ann}\,\Omega)$$ with
definition as follows
\begin{eqnarray}\label{15}
D^z(\nabla):=({\cal{G}},\omega,P_z\circ {\rm Tor})
\;\;\;\;,\;\;\;\; \forall \nabla\in{\cal{D}}(\Omega,\big<.,.\big>)
\end{eqnarray}
where ${\cal{G}}:=\nabla_zz$ and $\omega:=\frac{1}{2} {\rm rot}\,
z$, is a one to one correspondence.\\

\noindent{\it Proof:} Clearly, the map is well-defined. First, we
prove that the map is one to one. In the view of theorem 1 and the
definition of $A$ we have
\begin{eqnarray}
\Omega\circ {\rm Tor}&=&A(.,.)-d\Omega(.,.)z-[.,.] \label{16}\\
D^z(\tilde\nabla)&=&D^z(\nabla),\label{17}
\end{eqnarray}
so $\tilde{A} = A$, $\tilde{\omega} = \omega$, and
$\tilde{\cal{G}}={ \cal{G}}$. By (\ref{5}), we give
\begin{eqnarray*}
\big<P_z(\tilde\nabla_XY)-P_z(\nabla_XY,V)\big>=0,
\end{eqnarray*}
and therefore, we have
\begin{eqnarray}
\tilde\nabla_XY-\nabla_XY=P_z(\tilde\nabla_XY-P_z(\nabla_XY))=0\label{17}
\end{eqnarray}
and so the map is one to one. Furthermore, $D^z$ is surjective,
since if we assume that $\cal{G}\in{\rm Ann}\Omega$,
$\omega\in\wedge^2({\rm Ann}(\Omega))$, and
$\Theta\in\wedge^2(TM,{\rm Ann}(\Omega))$ be fixed, then by
(\ref{15}) for arbitrary $X,Y\in\Gamma(TM)$ we have
\begin{eqnarray*}
A(X,Y)=\Theta(X,Y)+d\Omega(X,Y)z+[X,Y].
\end{eqnarray*}
Hence, the following relation exist
\begin{eqnarray*}
\Omega(A(X,Y)) &=& d\Omega(X,Y)+\Omega([X,Y]) \\
&=& X(\Omega(Y))-Y(\Omega(X))
\end{eqnarray*}
and so for any $W,W_1,W_2\in\Gamma(\mathop{Ann}\Omega)$ we have
$$A(z,W)\in\Gamma(\mathop{\rm Ann}\Omega)\;\;\;,\;\;\;A(W_1,W_2)\in\Gamma(\mathop{\rm Ann}\Omega).$$
As a result, there is a unique map
$$\Pi:\Gamma(TM)\times\Gamma(TM\longrightarrow\Gamma(TM))$$
such that for any $X,Y\in\Gamma(TM)$ and $V\in\Gamma({\rm
Ann}\Omega)$, then $2\Big<\Pi(X,Y),V\Big>$ for fixed $\cal{G}$,
$\omega$, and $A$ that we assumed before, satisfies in relation
(\ref{5}).\par Now, for every $X,Y\in\Gamma(TM)$ we define
\begin{eqnarray*}
\nabla_XY=X(\Omega(Y))z+\Pi(X,Y),
\end{eqnarray*}
and by a straightforward computation, we can see that $\nabla_XY$
is a Galilean connection with
$D^z(\nabla)=(\cal{G},\omega,\Theta)$.\hfill $\diamond$
\end{document}